\documentclass[journal,twoside,web]{ieeecolor}
\usepackage{jsen}
\usepackage{cite}
\usepackage{amsmath,amssymb,amsfonts}
\usepackage{algorithmic}
\usepackage{graphicx}
\usepackage{textcomp}
\usepackage{wrapfig}
\def\BibTeX{{\rm B\kern-.05em{\sc i\kern-.025em b}\kern-.08em
    T\kern-.1667em\lower.7ex\hbox{E}\kern-.125emX}}
\definecolor{abstractbg}{rgb}{0.89804,0.94510,0.83137}
\begin{document}
\title{Lightweight Channel Codes for ISI Mitigation in Molecular Communication between Bionanosensors}
\author{Dongliang Jing, Andrew W. Eckford, \IEEEmembership{Senior Member, IEEE}
% \thanks{This paragraph of the first footnote will contain the date on 
% which you submitted your paper for review. It will also contain support 
% information, including sponsor and financial support acknowledgment. For 
% example, ``This work was supported in part by the U.S. Department of 
% Commerce under Grant BS123456.'' }
\thanks{Dongliang Jing is with the College of Mechanical and Electronic Engineering, Northwest A\&F University, China, Yangling, 712100 (e-mail: dljing@nwafu.edu.cn). }
\thanks{Andrew W. Eckford is with the department of Electrical Engineering and Computer Science, York University, Toronto, ON, Canada (e-mail: aeckford@yorku.ca).}}

\IEEEtitleabstractindextext{%
\fcolorbox{abstractbg}{abstractbg}{%
\begin{minipage}{\textwidth}%
\begin{abstract}
Channel memory and inter-symbol interference (ISI) are harmful factors in diffusion-based molecular communication (DBMC) between bionanosensors. To tackle these problems, this paper proposes a lightweight ISI-mitigating coding scheme to improve the system performance by shaping the signal using a constrained code. To characterize the proposed coding scheme theoretically, we derive analytical expressions for the bit error rate (BER) and the achievable rate based on Central Limit Theorem. Computer simulations are conducted to verify the accuracy of the theoretical results and demonstrate the superiority of the proposed coding scheme compared with the existing coding schemes.
\end{abstract}

\end{minipage}}}

\maketitle

\section{Introduction}
\label{sec:introduction}
\IEEEPARstart{W}{ith} 
the rapid development of nanotechnology, bionanosensors have been applied in many fields such as biological detection, intelligent drug delivery, and environmental monitoring. During the COVID-19 pandemic, bionanosensors were proposed to detect and model the spread of infections and diseases \cite{chen2021resource},\cite{khalid2020modeling}. 
However, for nanoscale applications, the constraint of the ratio of antenna size to electromagnetic signal wavelength makes the application of electromagnetic communication between bionanosensors infeasible. Inspired by communication in nature, a feasible way to solve this problem is molecular communication (MC), in which chemical signals are employed as the carrier of information during the information transmission.

A bacteria-based bionanosensor receiver is considered in \cite{bicen2019statistical}, which can be employed to connect the engineered biological system and the outside world. Diffusion-based molecular communication (DBMC) is a short-to-medium range MC without external energy and infrastructure\cite{yuan2018performance}, therefore, achieves a lot of research. In DBMC, concentration-based and molecular type-based modulation schemes were proposed in \cite{mahfuz2010characterization} and \cite{kuran2011modulation}, respectively. In \cite{li2019novel} and \cite{li2019asymmetric}, a time-based modulation scheme is proposed where information is encoded in the release time of molecules. In \cite{lin2015diffusion} and \cite{lin2016synchronization}, clock synchronization schemes were studied for DBMC. A communications scheme with asynchronous diffusion-based modulation is proposed in \cite{li2019clock}, which enables clock-free operation at the receiver.

In DBMC, the released molecules in the medium follows stochastic trajectories, described as Brownian motion \cite{arjmandi2017isi}. The Brownian motion of released molecules can degrade the performance of MC systems, as molecules arrive at random times and out of order. This leads to severe inter-symbol interference (ISI) problems and hinders communication at a high data rate \cite{gursoy2019pulse}.
%The released molecules may not arrive at the receiver in order leading to severe inter-symbol interference (ISI) problems and hinders communicating at high data rate, which degrades the performance of the MC systems \cite{gursoy2019pulse}.
Moreover, while classical channel equalization techniques rely on negative signal values to cancel ISI, MC systems cannot use these techniques because of the physical limitation that we can only release a positive amount of molecules into the medium \cite{mosayebi2018type}. Moreover, the size and power constraint of bionanosensors restrict the processing of complex tasks.

Previous studies have almost exclusively focused on the ISI effects in DBMC.
Accordingly, numerous potential approaches have been proposed for mitigating ISI, including modulation schemes, enzymes deployment, and channel coding schemes.
The modulation schemes have emphasized increasing the symbol duration or releasing similar type of molecules at sufficiently far apart time instances to mitigate ISI; this ensures that molecules of a previous transmission are degraded in the environment before molecules of the same type are reused for signaling.
In an attempt to reduce ISI effects, the work presented in \cite{tepekule2014novel} employed multiple molecule types simultaneously, which also supported in achieving a better symbol error rate. Other work presented the depleted molecule shift keying (D-MoSK) scheme, in which bits 1 are encoded onto different types of molecules and no molecule is released for bit 0 \cite{kabir2015d}. In \cite{arjmandi2017isi}, the authors have driven the further improvement of ISI mitigation scheme by adjusting the release time of the same type of molecules and by restricting types of molecules that can be used for encoding (i.e., reduced types of molecules as compared to MoSK). 
In \cite{gursoy2021concentration}, a hybrid modulation scheme which based on the pulse position and concentration of the released molecules is proposed to mitigate ISI.
The modification in the molecule types and their release time supported in reducing bit error rate (BER); however, it gives rise to delay and increases resource consumption.

Channel coding schemes involve adding additional bits to the input codewords for the sake of mitigating ISI in DBMC \cite{lu2015comparison, dissanayake2017reed, shih2012channel}. Similar to the traditional communication scheme, the channel codes address the errors instigated by ISI or channel noise. In order to improve the transmission reliability in a DBMC system, several channel coding schemes are employed in the literatures such as Euclidean geometry low density parity check (EG-LDPC) (LDPC), cyclic Reed-Muller (C-RM) codes \cite{lu2015comparison} and Reed-Solomon (RS) codes \cite{dissanayake2017reed}.
However, in \cite{shih2012channel}, it was shown that the performance of certain convolutional codes in the DBMC is worse than uncoded transmission. Therefore, channel coding schemes must be tailored for the DBMC instead of simply using an off-the-shelf code.

In this paper, a novel lightweight channel coding scheme called an ISI-mitigating code is proposed that studies the resource constraint nature of bionanosensors explicitly and improves the reliability of communication between bionanosensors. Our code is a constrained code with 
no consecutive bit 1 within the codeword, and no bit 1 at the last bit of codeword. Thus, there are no crossovers from the contiguous codewords. Moreover, it can be implemented with very low computational complexity, making it especially useful for nanoscale molecular communication applications. 
In spite of its simplicity, we show that the ISI-mitigating code preserves the BER performance with the increase of considering memory length.

The main contribution of the proposed work is a lightweight coding scheme that reduces ISI and is appropriate for molecular communication.
In the proposed scheme, consecutive bits 1 are never placed in the codeword, and the final codeword bit is always zero. The proposed scheme assists in avoiding the effect of molecular accumulation on the codeword, as well as supporting the correction of errors in the codeword.
Another key contribution of the paper is our analysis of this simplified scheme, deriving analytical expressions for the BER and the achievable rate for our constrained code considering a length-one channel memory. However, our results are not confined to length-one channel memory, and we present simulation results showing good performance for much longer memory, as is often found in molecular communication systems.

The remainder of this paper is organized as follows. Section \uppercase\expandafter{\romannumeral 2} describes the considered system model. In section \uppercase\expandafter{\romannumeral 3}, the ISI-mitigating coding scheme and analytical results are discussed. The expression for BER and the achievable rate of the proposed DBMC system is presented in \uppercase\expandafter{\romannumeral 4}. The performance of the proposed channel coding scheme is evaluated in
\uppercase\expandafter{\romannumeral 5}, and finally, conclusions are given in Section \uppercase\expandafter{\romannumeral 6}.

\section{System Model}
In this section, we employ the Brownian motion channel model to describe the diffusion process of releasing the molecules.
The overall system with a transmitter, medium, and a receiver is shown in Fig. 1. The radius of the sphere receiver is $r$, and the distance between the transmitter and the surface of the receiver is $d$. For the sake of simplicity, the following assumptions are used to describe the system:

\begin{enumerate}
\item The transmitter is a point source and is the only source of molecules.

\item The motion of released molecules is independent and identically distributed (iid), with a constant diffusion coefficient $D$.

\item The medium is infinite in every direction, without barrier or obstacle except the receiver.

\item The receiver is an absorbing sphere with the ability to count and absorb the received molecules.
\end{enumerate}

The transmitter emits a different number of molecules to represent each symbol, i.e., concentration shift keying (CSK) \cite{kuran2011modulation}.
The receiver demodulates the signals based on the following rules:

\begin{equation}
\small
{\rm{The{\kern 4pt} received{\kern 4pt} symbol{\kern 4pt}}}
= \left\{ {\begin{array}{*{20}{c}}
{1,}&{{N_r} \ge {\tau }}\\
{0,}&{{N_r} < {\tau }}
\end{array}\begin{array}{*{20}{c}},\\
\end{array}} \right.
\end{equation}
where $N_r$ denotes the number of received molecules, and $\tau$ is the decision threshold.
 \begin{figure}[!t]
  \centering
  \includegraphics[width=0.45\textwidth]{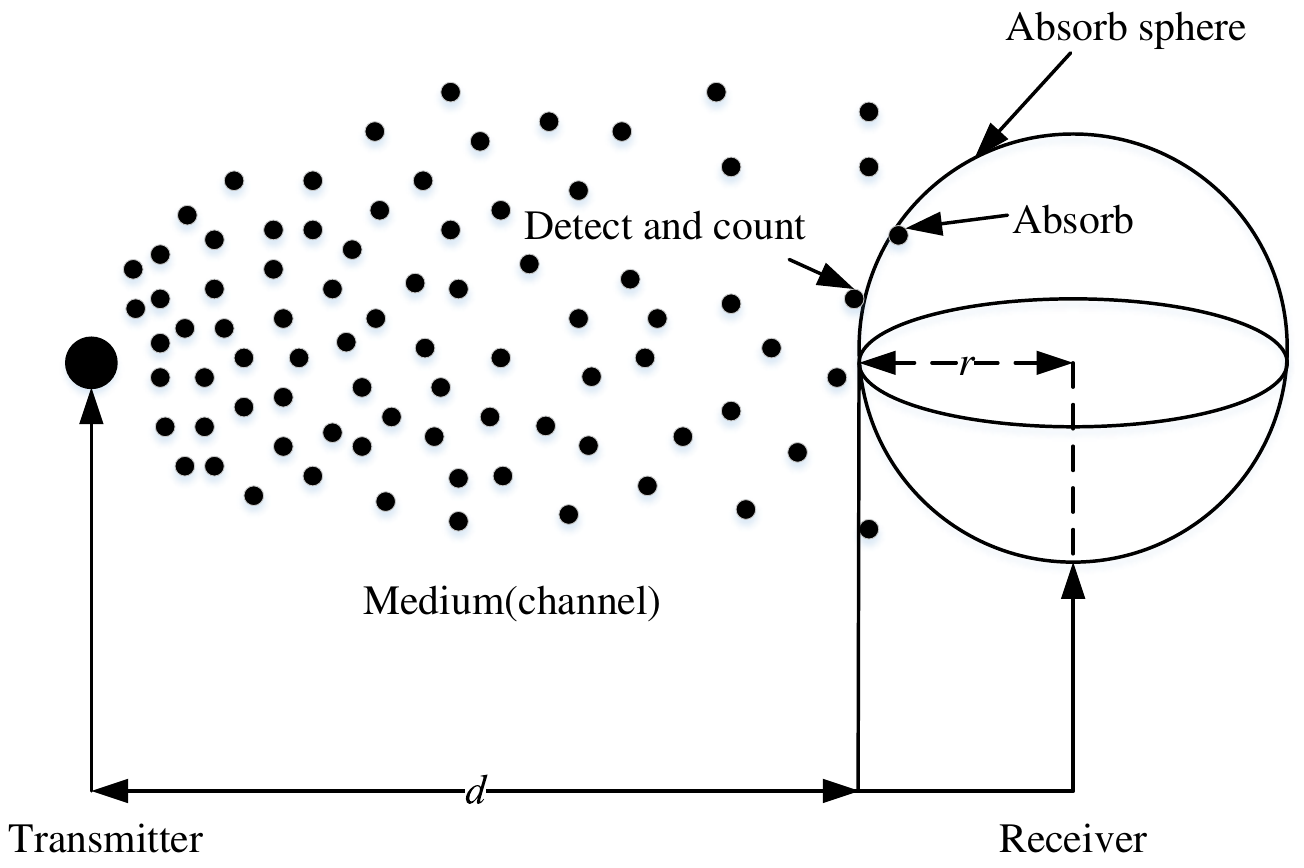}\\
  \caption{The considered molecular communication system.}
\end{figure}

All the molecules are absorbed as soon as they arrive at the receiver, in other words, the molecules account for the communication only one time.

\begin{figure}[!t]
  \centering
  \includegraphics[width=0.45\textwidth]{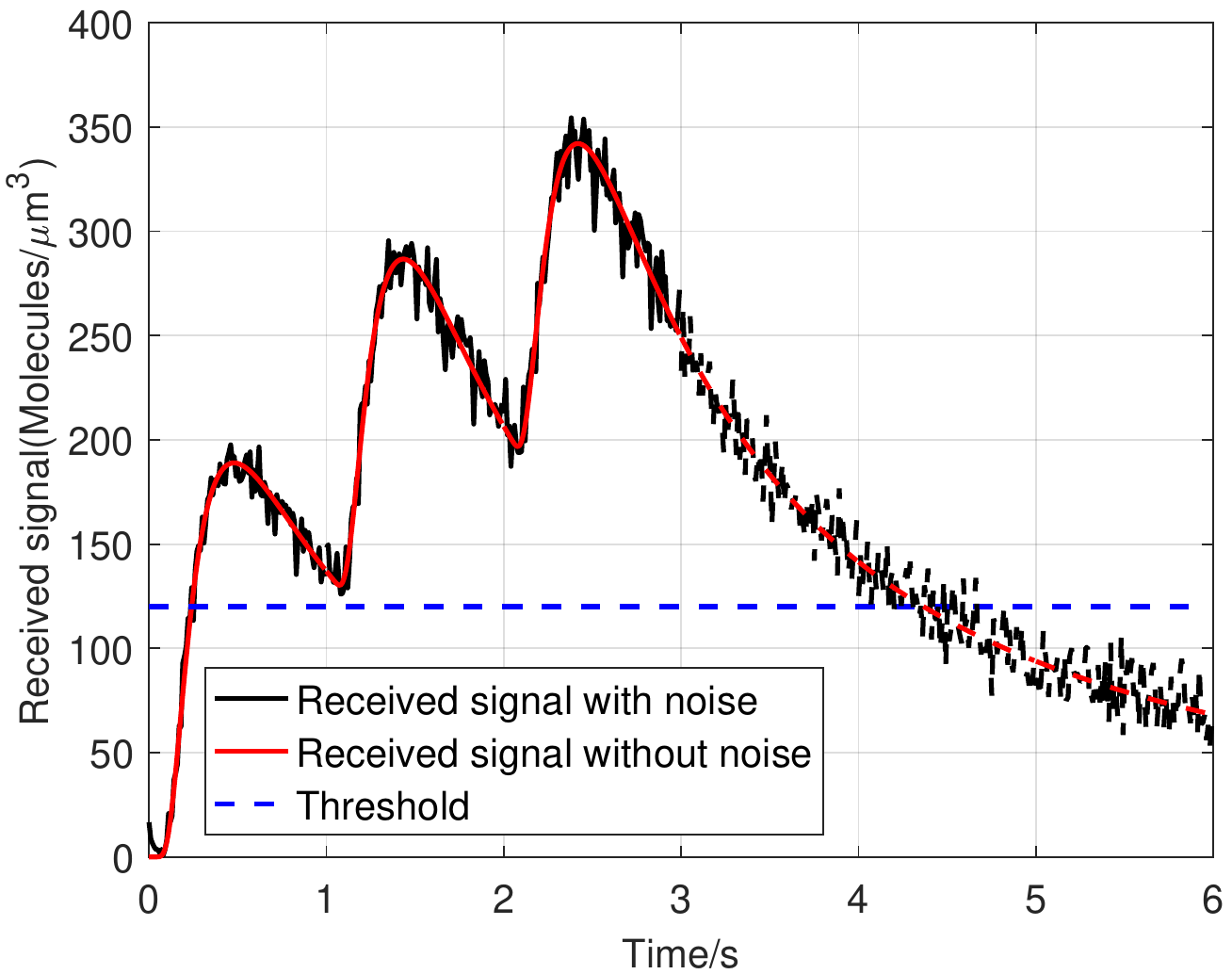}\\
  \caption{The received signal with consecutive bits 1 transmitted.} \label{diffusion}
\end{figure}

\begin{figure}[!t]
  \centering
  \includegraphics[width=0.45\textwidth]{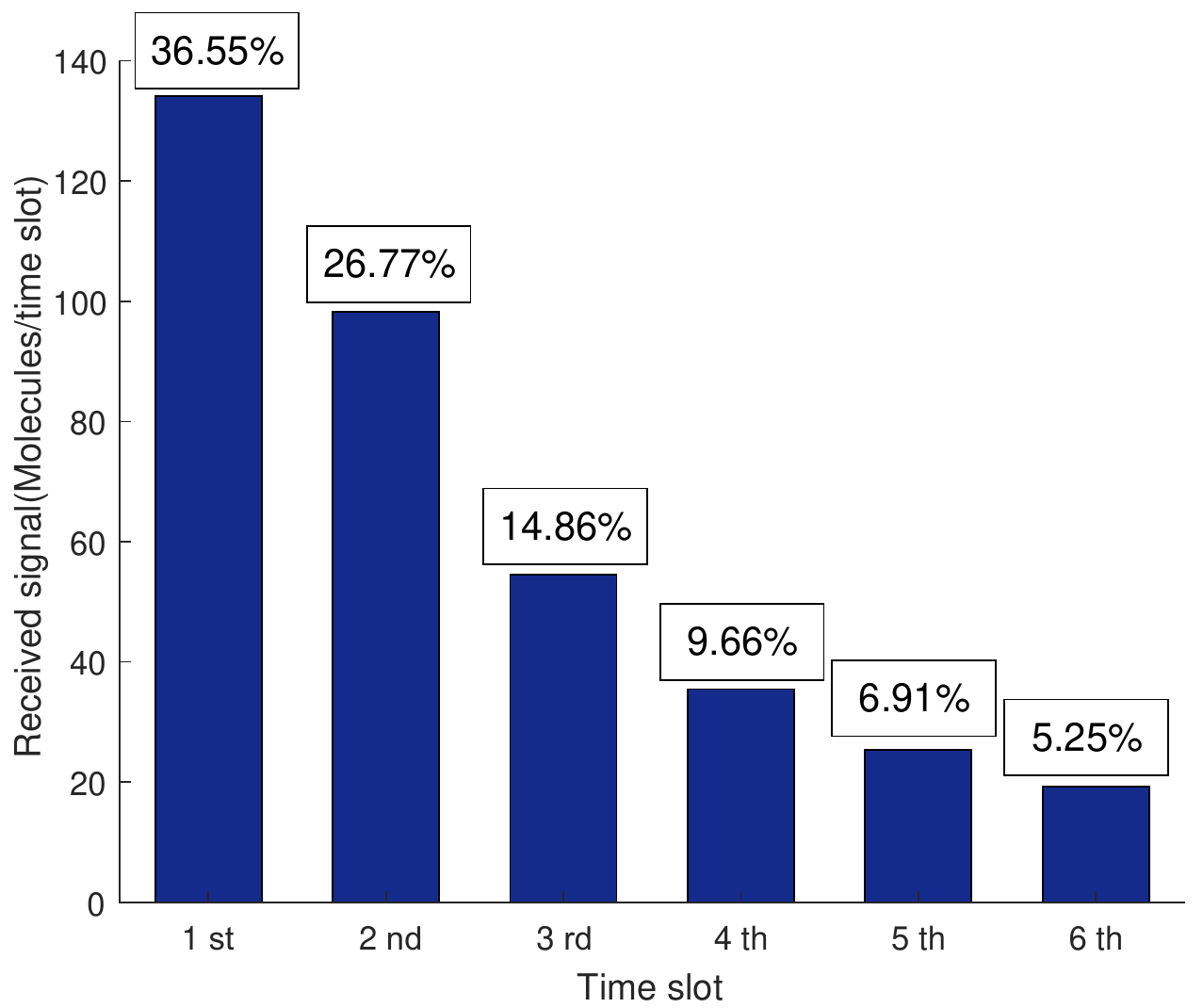}\\
  \caption{The molecules accumulate at the receiver after transmission of bit 1.}\label{number}
\end{figure}

The released molecules diffuse freely in the medium, via Brownian motion. In the diffusion process, the molecules follow different trajectories, and only part of the molecules are absorbed by the receiver, as shown in Fig. \ref{diffusion}. The remaining molecules may arrive at the receiver in the later time slots, resulting in the memory of the channel and ISI. As can be easily seen from Fig. \ref{number}, the remaining molecules mainly influence the following time slot and the total number of received molecules in the current and the following time slots is more than $63\%$. In the considered DBMC system, we select  $T_b = 2T_{peak}$, where $T_b$ is the symbol duration and $T_{peak}$ is the time where the peak concentration is achieved. Then the most prominent effect on ISI is contributed by the nearest past bit $x_{i-1}$.
Therefore, in the performance analysis, we mainly considering the molecules from the current and the nearest previous time slot.

The diffusion of molecules can be described by Fick's second law shown in \cite{farsad2016comprehensive} as
\begin{equation}
\small
\frac{{\partial c(z,t)}}{{\partial t}} = {D}{\nabla ^2}c(z,t),
\end{equation}
where $c(z,t)$ is the average spatiotemporal concentration of molecules at the location $z$ and at the time $t$, $D$ is the diffusion coefficient. The probability density function of the time for a molecule arriving at the receiver is called first hitting time, that can be written as
\begin{equation}
\small
{f_d}\left( t \right) = \left\{ {\begin{array}{*{20}{c}}
0&{t \le 0}\\
{\frac{d}{{\sqrt {4\pi D{t^3}} }}\exp \left( { - \frac{{{d^2}}}{{4Dt}}} \right)}&{t > 0}.
\end{array}} \right.
\end{equation}

Specifically, the hitting rate of molecules on the receiver is given by \cite{yilmaz2014three}
\begin{equation}
\small
{f_d}\left( t \right) = \left\{ {\begin{array}{*{20}{c}}
0&{t \le 0}\\
{\frac{r}{{r + d}}\frac{d}{{\sqrt {4\pi D{t^3}} }}\exp \left( { - \frac{{{d^2}}}{{4Dt}}} \right)}&{t > 0}.
\end{array}} \right.
\end{equation}

The fraction of molecules absorbed by the receiver during $0$ to $t$ is

\begin{equation}
\small
{p_d}\left( t \right) = \int_0^t {{f_d}\left( t \right)} dt = \frac{r}{{r + d}}\text{erfc}\left( {\frac{d}{{\sqrt {4Dt} }}} \right).
\end{equation}

Hence, the expected number of molecules absorbed by the receiver in a time slot $[(i-1)T_b, iT_b]$ after $N_m$ molecules are released can be calculated by
\begin{equation}
\small
E\left[ {{N_r}\left( {(i-1)T_b, iT_b} \right)} \right] = {N_m}\left[ {{p_{d}}\left( {iT_b} \right) - {p_{d}}\left( (i-1)T_b \right)} \right],
\end{equation}
where $N_r$ denotes the number of received molecules.

In DBMC, the number of absorbed molecules during a time slot $[(i-1)T_b, iT_b]$ after $N_m$ molecules are released can be modeled as binomial distribution\cite{dissanayake2017reed}. It is known that, the binomial distribution can be approximated using the normal distribution, when $N_m$ is large, and ${p_d}(t)$ is not near 0 or 1. Irrespective of ISI, the distribution of absorbed molecules during $[(i-1)T_b, iT_b]$ due to the molecules released at the beginning of $[(i-1)T_b, iT_b]$ can be expressed as
\begin{equation}
\small
{N_{r,i}} \sim \mathcal{N}\left( {{N_m}{p_{d}}\left( T_b \right),{N_m}{p_{d}}\left( T_b \right)\left( {1 - {p_{d}}\left( T_b \right)} \right)} \right).
\end{equation}

Due to the channel memory, the number of molecules absorbed during $T_b$, account for the number of molecules received from the current bit ($x_i$) and all the previous bits ($x_{1:i-1}$) can be expressed as
\begin{equation}
\small
{N_{r,i,\rm{tot}}} = {N_{r,{i}}} + \sum\limits_{j = 1}^{i - 1} {{N_{r,{j}}}} + N_n,
\end{equation}
where ${N_{r,{i}}}$ is the number of molecules released at the beginning of $i$th bit interval and received during the $i$th bit interval and can be expressed as $N_{r,{i}}=x_iN_mp_{d,1}$, where $p_{d,1}$ is the fraction of received molecules which released at the beginning of $i$th bit interval and received during the $i$th bit interval; $N_{r,{j}}$ is the number of molecules released at the beginning of $j$th bit interval while received during the current bit interval (i.e., ISI) and can be expressed as $N_{r,{j}}= x_jN_m {{p_{d,i - j+1}}}$, and $N_n$ is the counting noise. The distributions of $N_{r,{i}}$, $N_{r,{j}}$, and $N_n$ can be expressed as 
\begin{align}
\small
\begin{split}
{N_{r,i}} \sim  {\cal N}\left( x_i{{N_m}{p_{d,1}},x_i{N_m}{p_{d,1}}(1 - {p_{d,1}})} \right)\\
{N_{r,{j}}} \sim  {\cal N}\left( x_j{{N_m}{p_{d,i-j+1}},x_j{N_m}{p_{d,i-j+1}}(1 - {p_{d,i-j+1}})} \right)
\end{split},
\end{align}

As for noise in DBMC, we assume that $N_n$ is additive white Gaussian noise (AWGN), which has been widely used in\cite{zhai2018anti, huang2019spatial, farsad2018neural} and demonstrate in \cite{farsad2014channel},
\begin{equation}
\small
\begin{array}{l}
{N_n} \sim  {\cal N}\left( 0, \sigma^2 \right) \vspace{5 pt} \\
\end{array},
\end{equation}
where $\sigma^2$ dependents on the expected number of molecules received by the receiver.

Considering the fact that most prominent effect on ISI is contributed by the nearest past bit $x_{i-1}$. Thus, considering ISI, the $N_{r,i,\rm{tot}}$ can be approximated as
\begin{equation}
\small
{N_{r,i,\rm{tot}}} = {N_{r,{i}}} + {N_{r,{{i - 1}}}} + N_n.
\end{equation}

\section {ISI-mitigating code}
In this section, we briefly introduce ISI-free codes and repetition codes. ISI-free code considers crossovers between the contiguous codewords as well as crossovers present within the codeword. Inspired by the ISI-free code, an ISI-mitigating code is proposed that mainly considers the ISI within the codeword. Later, we present the encoding and decoding of the ISI-mitigating code in detail.

\subsection{ISI-free code}
For a level-$L$ ISI-free code, the decoded information is error-free if all the crossovers are no more than $L$ lengths channel memory. Level-$L$ channel memory considers that the past $L$ bits affect the transmission of the current bit. The crossovers considered in \cite{shih2012channel} are divided into two parts: the crossovers between the contiguous codewords and the crossovers within the same codeword. To handle the first crossovers, the ISI-free coding scheme assigns two codewords to each input information bits, one starting with 0 and the other starting with 1.
The transmitter selects one from the two codewords based on the previous codeword. For the second kind of codewords, in order to avoid the decoding error, the level-$L$ crossover permutation sets from all different codewords are disjoint.

The codewords assignment of the ISI-free (4,2,1) code is shown in Table \uppercase\expandafter{\romannumeral 1}.

\subsection{Repetition code}
Repetition code is one of the common error-correcting codes in conventional communications. For the repetition code, the transmitter repeats the message odd times when sending the message; the receiver recovers the original message in the data stream that occurs most often. The repetition code is low in complexity but also low in code rate. In this paper, we select the repeat times $n =3$ to compare with the proposed ISI-mitigating code, as their code rate is close. We denote $n =3$ repetition code as repetition-3 code. The codeword assignment of the repetition-3 code is shown in Table \uppercase\expandafter{\romannumeral 2}.

\subsection{ISI-mitigating code}
Considering that DBMC suffers from ISI which is caused by the channel memory, leading to declining system performance. Based on the model described in Section \uppercase\expandafter{\romannumeral 2}, we propose the ISI-mitigating code, which is immune to level-$L$ channel memory.

A level-$L$ ISI-mitigating code guarantees part of the error decoded information can be corrected if the channel memory is no more than $L$ bits. The notation ($n, k$, $L$)  maps $k$-bit information to an $n$-bit codeword, and $L$ stands for the level-$L$ ISI-mitigating code.

In this article, we only consider the level-1 ISI-mitigating code in theory since level-1 channel memory happens more frequently than the higher level memory as shown in the Fig.2. That is, we only consider the nearest past bit $x_{i-1}$ affect the transmission of current bits. Higher-level ISI-mitigating code will be left for future work. The proposed ISI-mitigating (4, 2, 1) code with the codeword assignment is shown in Table \uppercase\expandafter{\romannumeral 3}.

The design of codeword is based on the following rules:

1. The complexity of the encoding and decoding is quite low, that can be fulfilled in the bionanosensors.

2. The bit 1 is placed at the front part of the codeword as far as possible to avoid the effect of contiguous codewords.

3. In the codeword, there is no consecutive bit 1 for the sake of avoiding the error decode, which is caused by accumulated molecules within the codeword.

Considering the Level-1 channel memory and noise, the receiver also probably error decodes the received bits. For example, \{0100\} may be error decoded as \{0010\} or \{0110\}. Based on the above encoding rules, the receiver is able to correct part of the error codes as shown in Table \uppercase\expandafter{\romannumeral 3}.
The analysis of the bit error rate is shown in section \uppercase\expandafter{\romannumeral 5}.

\begin{table}
\normalsize
\caption{The ISI-free (4,2,1)code.}
\centering
\begin{center}
\begin{tabular}{ | c | c | c | c |p{3cm}|}
\hline
Information bits & Starting with 0 & Starting with 1  \\ \hline
00 & 0000 & 1111 \\ \hline
01 & 0001 & 1000  \\ \hline
10 & 0011 & 1100 \\ \hline
11 & 0111 & 1110 \\ \hline
\end{tabular}
\end{center}
\end{table}

\begin{table}
\normalsize
\caption{The codeword of repetition-3 code.}
\centering
\begin{center}
\begin{tabular}{ | c | c | c | c |p{3cm}|}
\hline
Information bits & Encoded bits   \\ \hline
0 & 000   \\ \hline
1 & 111  \\ \hline
\end{tabular}
\end{center}
\end{table}

\begin{table}
\normalsize
\caption{The proposed Level-1 ISI-mitigating (4,2,1) codeword}
\centering
\begin{center}
\begin{tabular}{ | c | c | c | c |p{3cm}|}
\hline
Information bits & Encoded bits & Corrects error  \\ \hline
00 & 0000 &  \\ \hline
01 & 0100 & 0010  \\ \hline
10 & 1000 & 1100 \\ \hline
11 & 1010 & 1011, 1110, 1111 \\ \hline
\end{tabular}
\end{center}
\end{table}

\section{Performance analysis}
In this section, we analyze the code rate and complexity of the above three coding schemes. Furthermore, the bit error rate and achievable rate of the proposed ISI-mitigating code are also derived.

\subsection{Code rate}

For the Level-1 ISI-mitigating code and Level-1 ISI-free code, we can easily achieve the code rate: $E=n/k=1/2$.

For repetition-3 code, the code rate is :$E=1/n=1/3$.

As can be seen from above, the code rate of the ISI-mitigating code and ISI-free code are higher than that of repetition-3 code.

\subsection{Complexity analysis}
The time complexity is the time cost of information bit maps to the encoded bits. In this paper, we employ the maximum time complexity to express the time complexity. The space complexity is the temporarily occupying storage space during the encoding.
The time and space complexity of the above three channel coding schemes are shown in Table \uppercase\expandafter{\romannumeral 4}.
It can be seen from Table \uppercase\expandafter{\romannumeral 4}, the proposed ISI-mitigating code achieves lower time and space complexity.
\begin{table}
\normalsize
\caption{The complexity analysis of the three channel coding schemes}
\centering
\begin{center}
\begin{tabular}{ | c | c | c | c |p{3cm}|}
\hline
 & Time complexity & Space complexity  \\ \hline
ISI-free code   & $O(L*4)$ & $O(L*2)$\\ \hline
Repetition code & $O(L*2)$ & $O(L*3)$  \\ \hline
ISI-mitigating code & $O(L*2)$ & $O(L*2)$ \\ \hline
\end{tabular}
\end{center}
\end{table}

\subsection{Bit error rate}

To analyze the BER of the proposed channel code, we assume that the considered  channel noise $N_n<\tau$, which means the receiver would not decode the absorbed molecules as symbol 1 when there are no information molecules sent by the transmitter.  Specifically, for the uncoded binary channel, the conditional BER of the $i$th bit can be expressed as
\begin{equation}
\footnotesize
\begin{split}
\small
{P_e} = \mathrm{Pr}\left( {{y_i} = 0|{x_i} = 1,{x_{1:{{i - 1}}}}} \right)\sum\limits_{{x_k} \in \left\{ {0,1} \right\}} {\mathrm{Pr}\left( {{x_i} = 1} \right)\prod\limits_{k = 1}^{i - 1} {\mathrm{Pr}\left( {{x_k}} \right)} } \\
 + \mathrm{Pr}\left( {{y_i} = 1|{x_i} = 0,{x_{1:{{i - 1}}}}} \right)\sum\limits_{{x_k} \in \left\{ {0,1} \right\}} {\mathrm{Pr}\left( {{x_i} = 0} \right)\prod\limits_{k = 1}^{i - 1} {\mathrm{Pr}\left( {{x_k}} \right)} },
\end{split}
\end{equation}
where $x_k$ denotes the bit transmitted at the $k$th time slot. Considering $L$ length of channel memory, the received error can be expressed as
\begin{equation}
\small
\begin{split}
\displaystyle {P_{(e|{x_i} = 0,x_{i-L:i-1})}} = \frac{1}{{{2^{L+1}}}}\sum\limits_{{x_{i - L}}, \ldots {x_{i - 1}} \in \left( {0,1} \right)} {\mathrm{Pr}\left( {\sum\limits_{l = 0}^L {{N_{r,{i - l}}} \ge \tau } } \right)} ,\\
\displaystyle {P_{(e|{x_i} = 1,x_{i-L:i-1})}} = \frac{1}{{{2^{L+1}}}}\sum\limits_{{{i - L}}, \ldots {x_{i - 1}} \in \left( {0,1} \right)} {\mathrm{Pr}\left( {\sum\limits_{l = 0}^L {{N_{r,{i - l}}} < \tau } } \right)}.
\end{split}
\end{equation}
As discussed above, only considering the current and the nearest past bit, then, the BER can be expressed as
\begin{align}
\small
\begin{split}
\label{ber_average}
{P_e} = \mathrm{Pr} \left( x_{i}, x_{i-1} \right)\mathrm{Pr}\left( \mathrm{error} \:|\:  x_{i}, x_{i-1} \right) .
\end{split}    
\end{align}

An example is taken to explain the analysis of BER more clearly. To transmit bits 10, the transmitter encodes them as 1000. Considering one length of channel memory and the fact $N_n<\tau$, the decoded bits and error probability can be expressed as
\begin{align}
\small
10 {\kern -2pt}\to {\kern -2pt}1000 {\kern -2pt}\to {\kern -2pt}\left\{ {\begin{array}{*{1}{c}}
{{\kern -6pt}0000 {\kern -2pt}\to  \times  {\kern -2pt}\to {\kern -2pt}\mathrm{Pr}\left( {{N_{r,i}} {\kern -2pt}+ {\kern -2pt}{N_n} {\kern -2pt}< {\kern -2pt}\tau } \right)\mathrm{Pr}\left( {{N_{r,{i-1}}} {\kern -2pt}+{\kern -2pt} {N_n} {\kern -2pt}< {\kern -2pt}\tau } \right)}\\
{{\kern -6pt}0100 {\kern -2pt}\to  \times  {\kern -2pt}\to {\kern -2pt}\mathrm{Pr}\left( {{N_{r,i}}{\kern -2pt} + {\kern -2pt}{N_n} {\kern -2pt}<{\kern -2pt} \tau } \right)\mathrm{Pr}\left( {{N_{r,{i-1}}}{\kern -2pt} +{\kern -2pt} {N_n} {\kern -2pt}> {\kern -2pt}\tau } \right)}\\
{{\kern -100pt}1000 {\kern -2pt}\to 10 {\kern -2pt}\to{\kern 33pt} 0}\\
{{\kern -100pt}1100 {\kern -2pt}\to 10 {\kern -2pt}\to{\kern 33pt} 0}
\end{array}} \right.
\end{align}

Considering that ISI-mitigating code is able to correct parts of error codes, the probability of bit error rate can be expressed as
\begin{equation}
\small
\begin{array}{l}
\mathrm{Pr}\left( {{y_i} = 0|{x_i} = 1,{x_{i - 1}} = 0} \right)\\
 = \frac{1}{4}\mathrm{Pr}\left( {{N_{r,i}} + {N_n} < \tau } \right),
\end{array}
\end{equation}

\begin{equation}
\small
\begin{array}{l}
\mathrm{Pr}\left( {{y_i} = 0|{x_i} = 1,{x_{i - 1}} = 1} \right)\\
 = \frac{1}{4} \mathrm{Pr}\left( {{N_{r,i}} + {N_{r,{i - 1}}} + {N_n} < \tau } \right),
\end{array}
\end{equation}

\begin{equation}
\small
\begin{array}{l}
\mathrm{Pr}\left( {{y_i} = 1|{x_i} = 0,{x_{i - 1}} = 1} \right)\\
 = \frac{1}{4} \mathrm{Pr}\left( {{N_{r,{i - 1}}} + {N_n} > \tau } \right),
\end{array}
\end{equation}

\begin{equation}
\small
\mathrm{Pr}\left( {{y_i} = 0|{x_i} = 0,{x_{i - 1}} = 0} \right) = \frac{1}{4},
\end{equation}
and
\begin{equation}
\small
\mathrm{Pr}\left( {{y_i} = 1|{x_i} = 0,{x_{i - 1}} = 0} \right) = 0.
\end{equation}

From equations (9) - (10), the equations (16) - (18) can be expressed as
\begin{small}
\begin{align} \notag
& \mathrm{Pr}\left( {{N_{r,i}} + {N_n} < \tau } \right)\\
& = 1 - Q\left( {\frac{{\tau  - {N_m}{p_{d,1}}}}{{\sqrt {{N_m}{p_{d,1}}\left( {1 - {p_{d,1}}} \right) + {\delta ^2}} }}} \right),
\end{align}
\end{small}

\begin{small}
\begin{align} \notag
&\mathrm{Pr}\left( {{N_{r,i}} + {N_{r,{i - 1}}} + {N_n} < \tau } \right)\\
 &= 1 - Q\left( {\frac{{\tau  - {N_m}\left({p_{d,1}}+{p_{d,2}}\right)}}{{\sqrt {{N_m}{p_{d,1}}\left( {1 - {p_{d,1}}} \right) + {N_m}{p_{d,2}}\left( {1 - {p_{d,2}}} \right) + {\delta ^2}} }}} \right),
\end{align}
\end{small}

\begin{small}
\begin{align} \notag
&\mathrm{Pr}\left( {{N_{r,{i-1}}} + {N_n} > \tau } \right)\\
 &= Q\left( {\frac{{\tau  - {N_m}{p_{d,2}}}}{{\sqrt { {N_m}{p_{d,2}}\left( {1 - {p_{d,2}}} \right) + {\delta ^2}} }}} \right).  
\end{align}
\end{small}

\subsection{Achievable Rate}
We define the achievable rate $C$ that maximizes the mutual information $I(X;Y)$ between the transmitted symbol $X$ and the received symbol $Y$ as follows:
\begin{equation}
\small
C = \mathop {\max }\limits_{{f_X}\left( x \right)} I\left( {X;Y} \right),
\end{equation}

with
\begin{equation}
\small
I\left( {X;Y} \right) = H\left( Y \right) - H\left( {Y|X} \right).
\end{equation}

In the considered DBMC system, the ISI is mainly affected by the most recent bit $x_{i-1}$. Therefore, we consider length-one channel memory, namely ${x_i}$ and ${x_{i-1}}$, the entropy $H(Y)$ of $Y$ can be expressed as
\begin{equation}
H\left( Y \right) = \sum\limits_{y_i \in \left\{ {0,1} \right\}} {{P_{Y}}\left( y_i \right){{\log }_2}\frac{1}{{{P_Y}\left( y_i \right)}}},
\end{equation}
where
\begin{equation}
{P_Y}\left( y_i \right) = \sum\limits_{\scriptstyle{x_i} \in \left\{ {0,1} \right\}\hfill\atop
\scriptstyle{x_{i - 1}} \in \left\{ {0,1} \right\}\hfill} {{P_{Y|X}}\left( {y_i|{x_i},{x_{i - 1}}} \right){P_X}\left( {{x_i},{x_{i - 1}}} \right)}.
\end{equation}

The conditional entropy $H(Y|X)$ of $Y$ can be expressed as the equation (28) at the next page.
\newcounter{mytempeqncnt0}
\begin{figure*}[t]
%\normalsize
\setcounter{mytempeqncnt0}{\value{equation}}
\setcounter{equation}{27}
\begin{small}
\begin{align}
\begin{array}{l}
H\left( {Y|X} \right) = \sum\limits_{\scriptstyle{x_i} \in \left\{ {0,1} \right\}\hfill\atop
\scriptstyle{x_{i - 1}} \in \left\{ {0,1} \right\}\hfill} {\sum\limits_{{y_i} \in \left\{ {0,1} \right\}} {{P_{X,Y}}\left( {{y_i},{x_i},{x_{i - 1}}} \right){{\log }_2}} \frac{1}{{{P_{Y|X}}\left( {{y_i}|{x_i},{x_{i - 1}}} \right)}}} \\
 = \sum\limits_{\scriptstyle{x_i} \in \left\{ {0,1} \right\}\hfill\atop
\scriptstyle{x_{i - 1}} \in \left\{ {0,1} \right\}\hfill} {P\left( {{x_i},{x_{i - 1}}} \right)\sum\limits_{{y_i} \in \left\{ {0,1} \right\}} {{P_{X|Y}}\left( {{y_i}|{x_i},{x_{i - 1}}} \right){{\log }_2}} \frac{1}{{{P_{Y|X}}\left( {{y_i}|{x_i},{x_{i - 1}}} \right)}}}
\end{array},
\end{align}
\end{small}
\setcounter{equation}{28}
\hrulefill
\vspace*{0pt}
\end{figure*}

Considering equations (19) - (20), we assume the conditional probability ${P_{Y|X}}\left( {y_i|{x_i},{x_{i - 1}}} \right)$ as
\begin{equation}
\small
{P_{Y|X}}\left( {y_i|{x_i},{x_{i - 1}}} \right) = \left\{ {\begin{array}{*{20}{c}}
{\begin{array}{*{20}{c}}
{\begin{array}{*{20}{c}}
{\begin{array}{*{20}{c}}
1&{y_i = 0,{x_i} = 0,{x_{i - 1}} = 0}
\end{array}}\\
{\begin{array}{*{20}{c}}
0&{y_i = 1,{x_i} = 0,{x_{i - 1}} = 0}
\end{array}}\\
{\begin{array}{*{20}{c}}
{{p_\alpha }}&{y_i = 0,{x_i} = 0,{x_{i - 1}} = 1}
\end{array}}\\
{\begin{array}{*{20}{c}}
{{p_\beta }}&{y_i = 1,{x_i} = 0,{x_{i - 1}} = 1}
\end{array}}
\end{array}}\\
{\begin{array}{*{20}{c}}
{{p_\gamma }}&{y_i = 0,{x_i} = 1,{x_{i - 1}} = 0}
\end{array}}\\
{\begin{array}{*{20}{c}}
{{p_\eta }}&{y_i = 1,{x_i} = 1,{x_{i - 1}} = 0}
\end{array}}
\end{array}}\\
{\begin{array}{*{20}{c}}
{\begin{array}{*{20}{c}}
{{p_\lambda }}&{y_i = 0,{x_i} = 1,{x_{i - 1}} = 1}
\end{array}}\\
{\begin{array}{*{20}{c}}
{{p_\mu }}&{y_i = 1,{x_i} = 1,{x_{i - 1}} = 1}
\end{array}}
\end{array}},
\end{array}} \right.
\end{equation}

where \[0 \le \left\{ {{p_\alpha },{p_\beta },{p_\gamma },{p_\eta },{p_\lambda },{p_\mu }} \right\} \le 1,\]

and
\begin{equation}
\small
\begin{array}{l}
{p_\alpha } + {p_\beta } = 1\\
{p_\gamma } + {p_\eta } = 1\\
{p_\lambda } + {p_\mu } = 1
\end{array}.
\end{equation}

 As the transmission of $x_i$ and $x_{i-1}$ is independent and with the same probability to send 0 and 1, thus, we assume

\begin{equation}
{P_X}\left( {{x_i},{x_{i - 1}}} \right) = {P_X}\left( {{x_i}} \right){P_X}\left( {{x_{i - 1}}} \right) = {p_\xi }.
\end{equation}

In Equation (28), $\sum\limits_{{y_i} \in \left\{ {0,1} \right\}} {{P_{X|Y}}\left( {{y_i}|{x_i},{x_{i - 1}}} \right){{\log }_2}} \frac{1}{{{P_{Y|X}}\left( {{y_i}|{x_i},{x_{i - 1}}} \right)}}$ can be expanded to equation (32) in the next page.

\newcounter{mytempeqncnt1}
\begin{figure*}[t]
%\normalsize
\setcounter{mytempeqncnt1}{\value{equation}}
\setcounter{equation}{31}
\begin{small}
\begin{align}
\begin{array}{l}
\sum\limits_{{y_i} \in \left\{ {0,1} \right\}} {{P_{X,Y}}\left( {{y_i}|{x_i},{x_{i - 1}}} \right){{\log }_2}} \frac{1}{{{P_{Y|X}}\left( {{y_i}|{x_i},{x_{i - 1}}} \right)}}\\
 = {p_\alpha }{\log _2}\frac{1}{{{p_\alpha }}} + {p_\beta }{\log _2}\frac{1}{{{p_\beta }}} + {p_{_\gamma }}{\log _2}\frac{1}{{{p_{_\gamma }}}} + {p_{_\eta }}{\log _2}\frac{1}{{{p_\eta }}} + {p_{_\lambda }}{\log _2}\frac{1}{{{p_\lambda }}} + {p_{_\mu }}{\log _2}\frac{1}{{{p_\mu }}}\\
 = {p_\alpha }{\log _2}\frac{1}{{{p_\alpha }}} + \left( {1 - {p_\alpha }} \right){\log _2}\frac{1}{{1 - {p_\alpha }}} + {p_{_\gamma }}{\log _2}\frac{1}{{{p_{_\gamma }}}} + \left( {1 - {p_{_\gamma }}} \right){\log _2}\frac{1}{{1 - {p_{_\gamma }}}} + {p_{_\lambda }}{\log _2}\frac{1}{{{p_\lambda }}} + \left( {1 - {p_{_\lambda }}} \right){\log _2}\frac{1}{{1 - {p_\lambda }}}\\
% = f\left( {{p_\alpha }} \right) + f\left( {{p_{_\gamma }}} \right) + f\left( {{p_{_\lambda }}} \right)\\
\end{array},
\end{align}
\end{small}
\setcounter{equation}{32}
\hrulefill
\vspace*{0pt}
\end{figure*}

In equation(32), we define
\begin{equation}
f\left( {{p_x }} \right) = {p_x }{\log _2}\frac{1}{{{p_x }}} + \left( {1 - {p_x }} \right){\log _2}\frac{1}{{1 - {p_x }}}.\\
\end{equation}
Thus, the equation (32) can be expressed as
\begin{equation}
\begin{array}{l}
\small
\sum\limits_{{y_i} \in \left\{ {0,1} \right\}} {{P_{X,Y}}\left( {{y_i}|{x_i},{x_{i - 1}}} \right){{\log }_2}} \frac{1}{{{P_{Y|X}}\left( {{y_i}|{x_i},{x_{i - 1}}} \right)}}\\
= f\left( {{p_\alpha }} \right) + f\left( {{p_{_\gamma }}} \right) + f\left( {{p_{_\lambda }}} \right)\\
\end{array}.
\end{equation}
Therefore the conditional entropy $H(Y|X)$ of $Y$ can be expressed as equation (35) in the next page.

The entropy $H(Y)$ of $Y$ in the equation (25) can be expanded to the equation (36) in the next page.

\newcounter{mytempeqncnt2}
\begin{figure*}[t]
%\normalsize
\setcounter{mytempeqncnt2}{\value{equation}}
\setcounter{equation}{34}
\begin{small}
\begin{align}
\begin{array}{l}
H\left( {Y|X} \right) = \sum\limits_{\scriptstyle{x_i} \in \left\{ {0,1} \right\}\hfill\atop
\scriptstyle{x_{i - 1}} \in \left\{ {0,1} \right\}\hfill} {\sum\limits_{{y_i} \in \left\{ {0,1} \right\}} {{P_{X,Y}}\left( {{y_i},{x_i},{x_{i - 1}}} \right){{\log }_2}} \frac{1}{{{P_{Y|X}}\left( {{y_i}|{x_i},{x_{i - 1}}} \right)}}} \\
 = \sum\limits_{\scriptstyle{x_i} \in \left\{ {0,1} \right\}\hfill\atop
\scriptstyle{x_{i - 1}} \in \left\{ {0,1} \right\}\hfill} {P\left( {{x_i},{x_{i - 1}}} \right)\sum\limits_{{y_i} \in \left\{ {0,1} \right\}} {{P_{X|Y}}\left( {{y_i}|{x_i},{x_{i - 1}}} \right){{\log }_2}} \frac{1}{{{P_{Y|X}}\left( {{y_i}|{x_i},{x_{i - 1}}} \right)}}} \\
 = {p_\xi }\left( {f\left( {{p_\alpha }} \right) + f\left( {{p_\gamma }} \right) + f\left( {{p_\lambda }} \right)} \right)
\end{array},
\end{align}
\end{small}
\setcounter{equation}{35}
\hrulefill
\vspace*{0pt}
\end{figure*}

\newcounter{mytempeqncnt}
\begin{figure*}[t]
%\normalsize
\setcounter{mytempeqncnt}{\value{equation}}
\setcounter{equation}{35}
\begin{small}
\begin{align} \notag
H\left( Y \right) &= \sum\limits_{y \in \left\{ {0,1} \right\}} {{P_Y}\left( y \right){{\log }_2}\frac{1}{{{P_Y}\left( y \right)}}} = \sum\limits_{y \in \left\{ {0,1} \right\}} {\left( {\sum\limits_{\scriptstyle{x_i} \in \left\{ {0,1} \right\}\hfill\atop
\scriptstyle{x_{i - 1}} \in \left\{ {0,1} \right\}\hfill} {{p_\xi }{P_{Y|X}}\left( {y|{x_i},{x_{i - 1}}} \right){{\log }_2}\frac{1}{{\sum\limits_{\scriptstyle{x_i} \in \left\{ {0,1} \right\}\hfill\atop
\scriptstyle{x_{i - 1}} \in \left\{ {0,1} \right\}\hfill} {{p_\xi }{P_{Y|X}}\left( {y|{x_i},{x_{i - 1}}} \right)} }}} } \right)} \\ \notag
 &= \sum\limits_{\scriptstyle{x_i} \in \left\{ {0,1} \right\}\hfill\atop
\scriptstyle{x_{i - 1}} \in \left\{ {0,1} \right\}\hfill} {{p_\xi }{P_{Y|X}}\left( {{y_i} = 0|{x_i},{x_{i - 1}}} \right){{\log }_2}\frac{1}{{\sum\limits_{\scriptstyle{x_i} \in \left\{ {0,1} \right\}\hfill\atop
\scriptstyle{x_{i - 1}} \in \left\{ {0,1} \right\}\hfill} {{p_\xi }{P_{Y|X}}\left( {y|{x_i},{x_{i - 1}}} \right)} }}} \\ \notag
 &+ \sum\limits_{\scriptstyle{x_i} \in \left\{ {0,1} \right\}\hfill\atop
\scriptstyle{x_{i - 1}} \in \left\{ {0,1} \right\}\hfill} {{p_\xi }{P_{Y|X}}\left( {{y_i} = 1|{x_i},{x_{i - 1}}} \right){{\log }_2}\frac{1}{{\sum\limits_{\scriptstyle{x_i} \in \left\{ {0,1} \right\}\hfill\atop
\scriptstyle{x_{i - 1}} \in \left\{ {0,1} \right\}\hfill} {{p_\xi }{P_{Y|X}}\left( {y|{x_i},{x_{i - 1}}} \right)} }}} \\ \notag
 &= {p_\xi }\left( {1 + {p_\alpha } + {p_{_\gamma }} + {p_\lambda }} \right){\log _2}\left( {\frac{1}{{{p_\xi }\left( {1 + {p_\alpha } + {p_{_\gamma }} + {p_\lambda }} \right)}}} \right) + {p_\xi }\left( {{p_\beta } + {p_\eta } + {p_\mu }} \right){\log _2}\frac{1}{{{p_\xi }\left( {{p_\beta } + {p_\eta } + {p_\mu }} \right)}}\\
 &= f\left( {{p_\xi }\left( {3 - \left( {{p_\alpha } + {p_{_\gamma }} + {p_\lambda }} \right)} \right)} \right).
\end{align}
\end{small}
\setcounter{equation}{36}
\hrulefill
\vspace*{0pt}
\end{figure*}

From equation (24) - (35), we achieve the mutual information $I\left( {X;Y} \right)$:
\begin{align}\notag
I\left( {X;Y} \right) &= H\left( Y \right) - H\left( {Y|X} \right)\\
 &= f\left( {{p_\xi }\left( {3 - \left( {{p_\alpha } + {p_{_\gamma }} + {p_\lambda }} \right)} \right)} \right) \\ \notag
 &- {p_\xi }\left( {f\left( {{p_\alpha }} \right) + f\left( {{p_\gamma }} \right) + f\left( {{p_\lambda }} \right)} \right),
\end{align}
where
\begin{align} \notag
\small
&{p_\alpha } = \mathrm{Pr}\left( {{y_i} = 0|{x_i} = 0,{x_{i - 1}} = 1} \right)\\ \notag
 &= \frac{1}{4}\mathrm{Pr}\left( {{N_{r,{i - 1}}} + {N_n} < \tau } \right)\\
 &= \frac{1}{4}\left[1 - Q\left( {\frac{{\tau  -  {{N_m}{p_{d,2}}} }}{{\sqrt {{N_m}{p_{d,2}}\left( {1 - {p_{d,2}}} \right) + {\delta ^2}} }}} \right)\right],\\ \notag
&{p_\gamma} = \mathrm{Pr}\left( {{y_i} = 0|{x_i} = 1,{x_{i - 1}} = 0} \right)\\ \notag
& = \frac{1}{4}\mathrm{Pr}\left( {{N_{r,i}} + {N_n} < \tau } \right)\\
& = \frac{1}{4}\left[1 - Q\left( {\frac{{\tau  - {N_m}{p_{d,1}}}}{{\sqrt {{N_m}{p_{d,1}}\left( {1 - {p_{d,1}}} \right) + {\delta ^2}} }}} \right)\right],
\end{align}

\begin{align}
\begin{split}
&{p_\lambda} = \mathrm{Pr}\left( {{y_i} = 0|{x_i} = 1,{x_{i - 1}} = 1} \right)\\
 &= \frac{1}{4}\mathrm{Pr}\left( {{N_{r,i}} + {N_{r,{i - 1}}} + {N_n} < \tau } \right)\\
 &= \frac{1}{4}\left[1 - \right. \\
 &\left. Q\left( {\frac{{\tau  - \left({p_{d,1}}+{p_{d,2}}\right)}}{{\sqrt {{N_m}{p_{d,1}}\left( {1 - {p_{d,1}}} \right) + {N_m}{p_{d,2}}\left( {1 - {p_{d,2}}} \right) + {\delta ^2}} }}} \right)\right].
\end{split}
\end{align}
Thus, the achievable rate $C$ can be achieved.

\section{performance evaluation}

In this section, we present the performance evaluation of the proposed ISI-mitigating code under DBMC described in section \uppercase\expandafter{\romannumeral 2}. Information bits 0 and 1 are transmitted randomly with the same probability. The simulation parameters are listed in Table \uppercase\expandafter{\romannumeral 5}.

\begin{figure}[!t]
  \centering
  \includegraphics[width=0.45\textwidth]{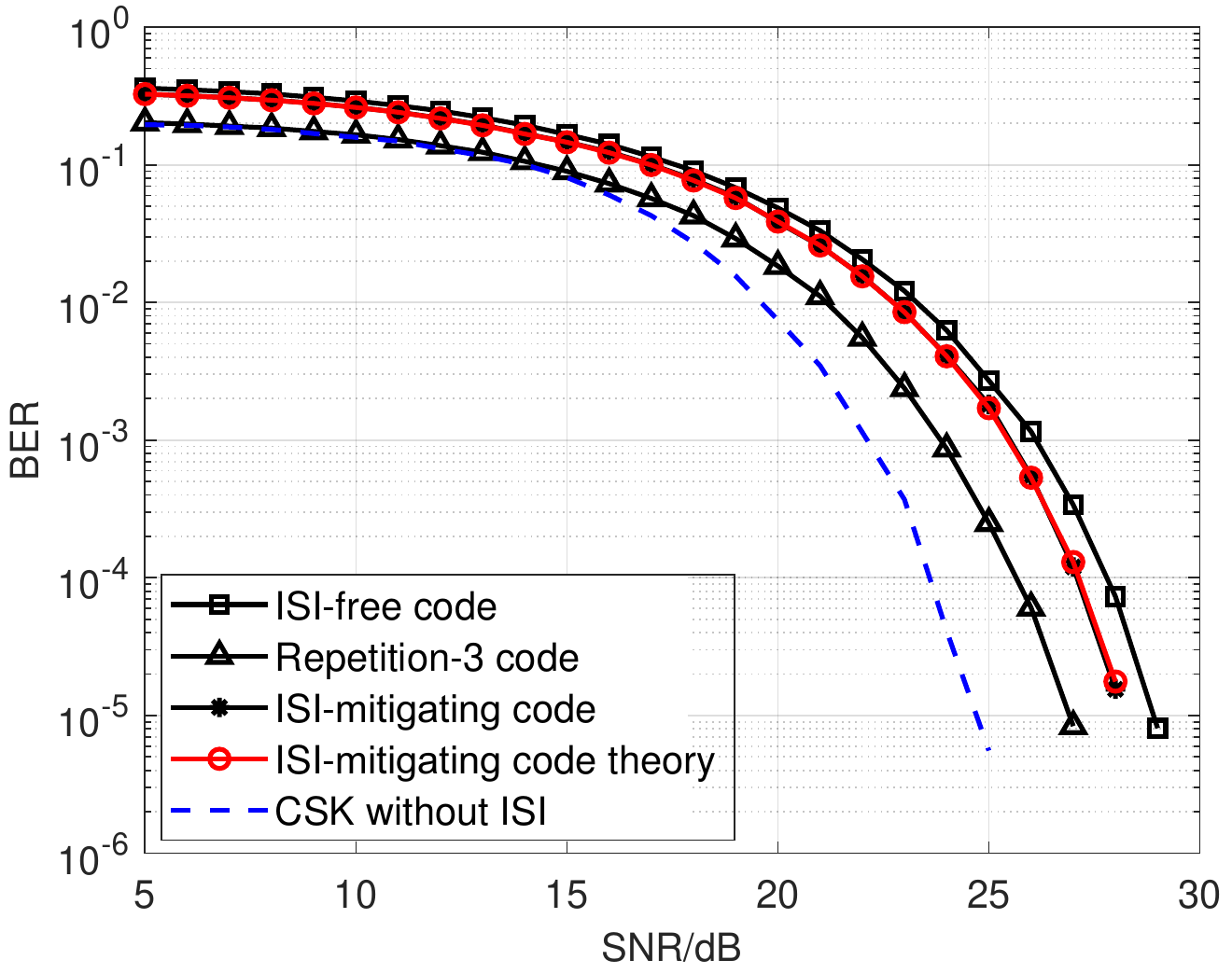}\\
  \caption{Performance comparison of the proposed ISI-mitigating code, ISI-free code, and repetition-3 code. The time slot $T_s=1s$ and the memory length $ L=1$.}\label{bervssnr}
\end{figure}

We first compare the performance of the proposed ISI-mitigating code, ISI-free code, and repetition-3 code under the memory length $L=1$. Performance comparisons among different codes are made with respect to the signal-to-noise ratio (SNR). The SNR, we use here, is defined as the ratio of the received signal power to the noise power\cite{kim2013novel}. From Fig. \ref{bervssnr} it is evident that the ISI-mitigating code outperforms the ISI-free code while the repetition-3 code achieves the best performance among all discussed schemes. This is because the ISI-mitigating code is able to correct some of the erroneous decoded bits, while the repetition-3 code is able to correct the erroneous decoded bits as long as the erroneous bits are not more than half of the transmitted bits. The repetition-3 code and the ISI-mitigating code are much closer to the CSK without ISI, verifying the effectiveness of the proposed ISI-mitigating code, in which BER performance is slightly inferior to the repetition-3 code but achieves higher code rate.
\begin{table}
\small
\caption{simulation parameters}
\centering
\begin{center}
\begin{tabular}{ | c | c | c | c |p{3cm}|}
\hline
Diffusion coefficient &$50 \times {10^{ - 5}}c{m^2}/s$  \\ \hline
Drift velocity & 0  \\ \hline
Receiver radius & $5 \times {10^{ - 5}}cm$  \\ \hline
Distance of Tx and Rx & $12 \times {10^{ - 5}}cm$  \\ \hline
\end{tabular}
\end{center}
\end{table}
Besides, the theoretical result of the ISI-mitigating code matches the simulation result very well, confirming the accuracy of the theoretical BER approximation.

\begin{figure}[!t]
  \centering
  \includegraphics[width=0.45\textwidth]{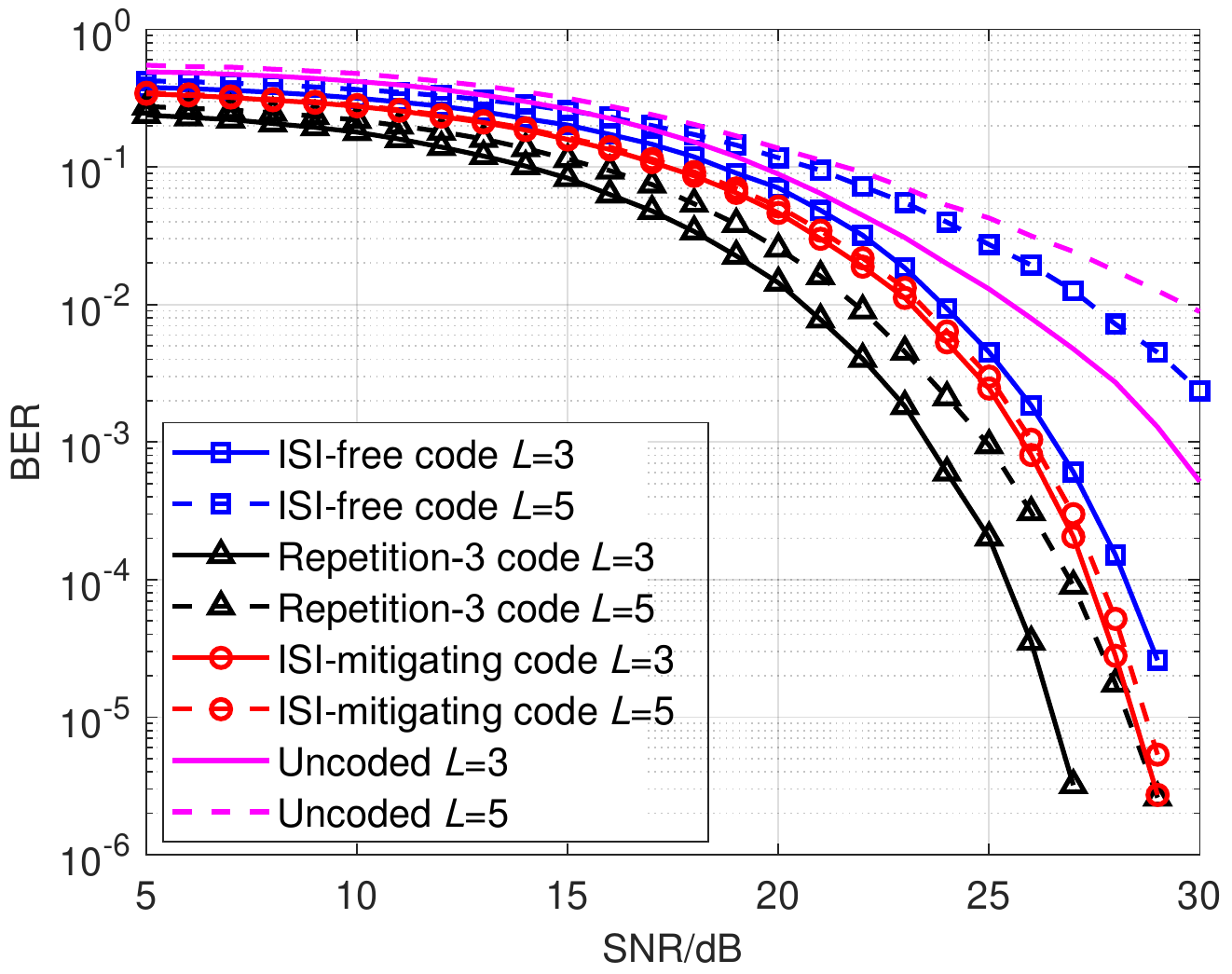}\\
  \caption{Comparison the BER of ISI-mitigating code, ISI-free code, and repetition-3 code under different memory length $L$.}\label{DMLL35}
\end{figure}

From the nature of diffusion and Fig. \ref{number}, we know only part of the released molecules are absorbed by the receiver at the current time slot. The remaining molecules will affect many subsequent time slots. Thus, good coding schemes should not sensitive to the memory length. From Fig. \ref{DMLL35}, we can see the BER of the repetition-3 code is obviously worse with the increase of memory length. However, the proposed ISI-mitigating code is almost unaffected by the memory length. This is because the proposed ISI-mitigating code is not only immune to Level-1 channel memory, which has maximum impact on the channel, but also resistant to longer memory length, as there is no consecutive bit 1 and fewer molecules in total compared with repetition-3 code, avoiding the accumulation of molecules in the channel. In particular, the repetition-3 code and ISI-mitigating code are much better than the uncoded communication system. Though the  ISI-mitigating code is slightly inferior to the repetition-3 code, however, the ISI-mitigating code achieves a better code rate.

\begin{figure}[!t]
  \centering
  \includegraphics[width=0.45\textwidth]{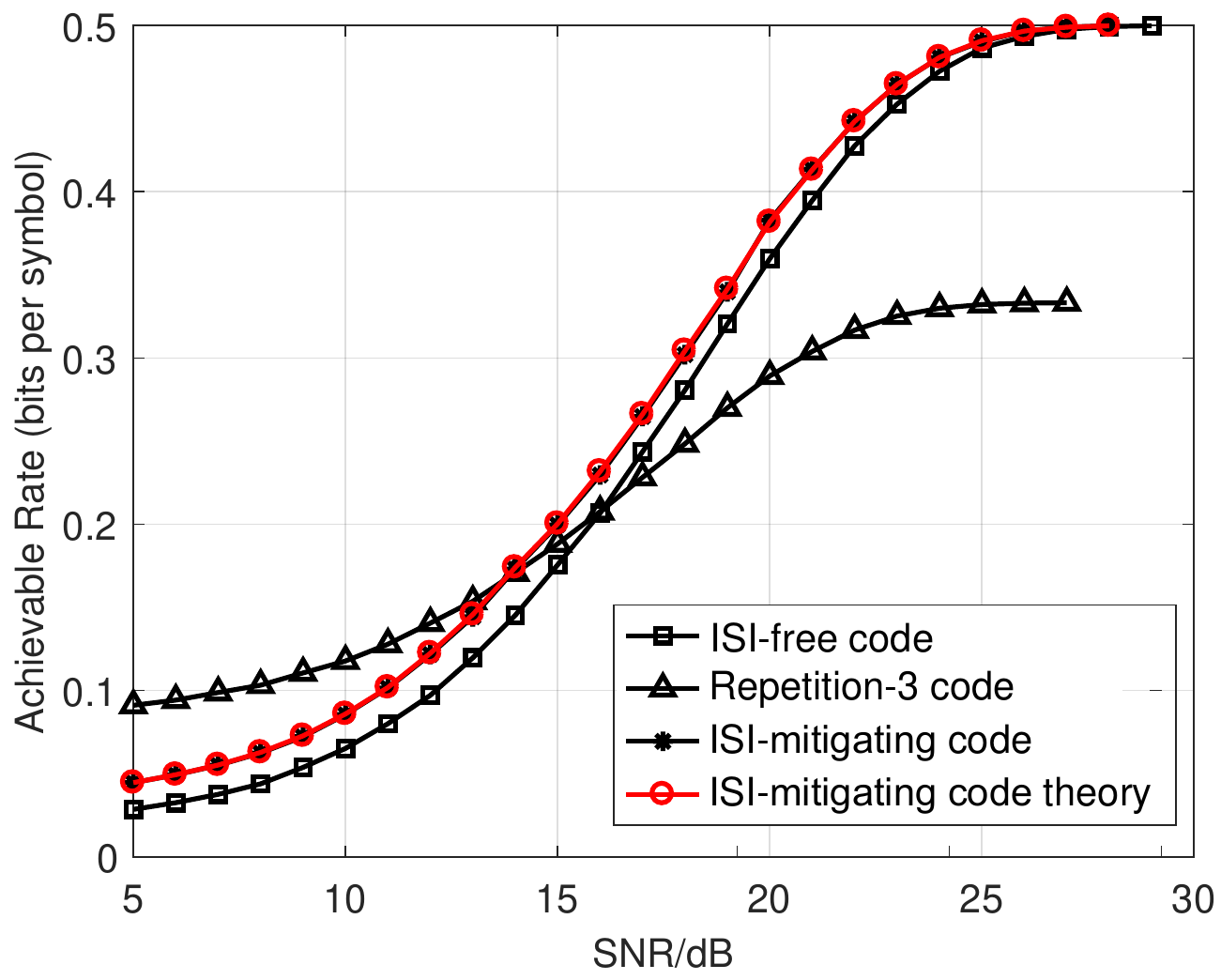}\\
  \caption{Achievable rate comparisons of different channel codes.}\label{achiveablerate}
\end{figure}
{\color{black}Fig. \ref{achiveablerate} shows the achievable rates per bit of ISI-mitigating} code, ISI-free code, and repetition-3 code. Fig. \ref{achiveablerate} depicts that, the ISI-mitigating code achieves the maximum achievable rate, and ISI-free code is inferior to the ISI-mitigating code but much better than the repetition-3 code. This is mainly because the code rate of ISI-mitigating code and ISI-free code is $1/2$ while the repetition-3 code is $1/3$. When SNR$\geq$16, achievable rates of the ISI-mitigating code and ISI-free are larger than the repetition-3 code. From this result, we can conclude that the repetition-3 code can be selected for high communication reliability while low achievable rate MC system. Meanwhile, the ISI-mitigating code achieves a balance between the communication reliability and achievable rate.

\begin{figure}[!h]
  \centering
  \includegraphics[width=0.45\textwidth]{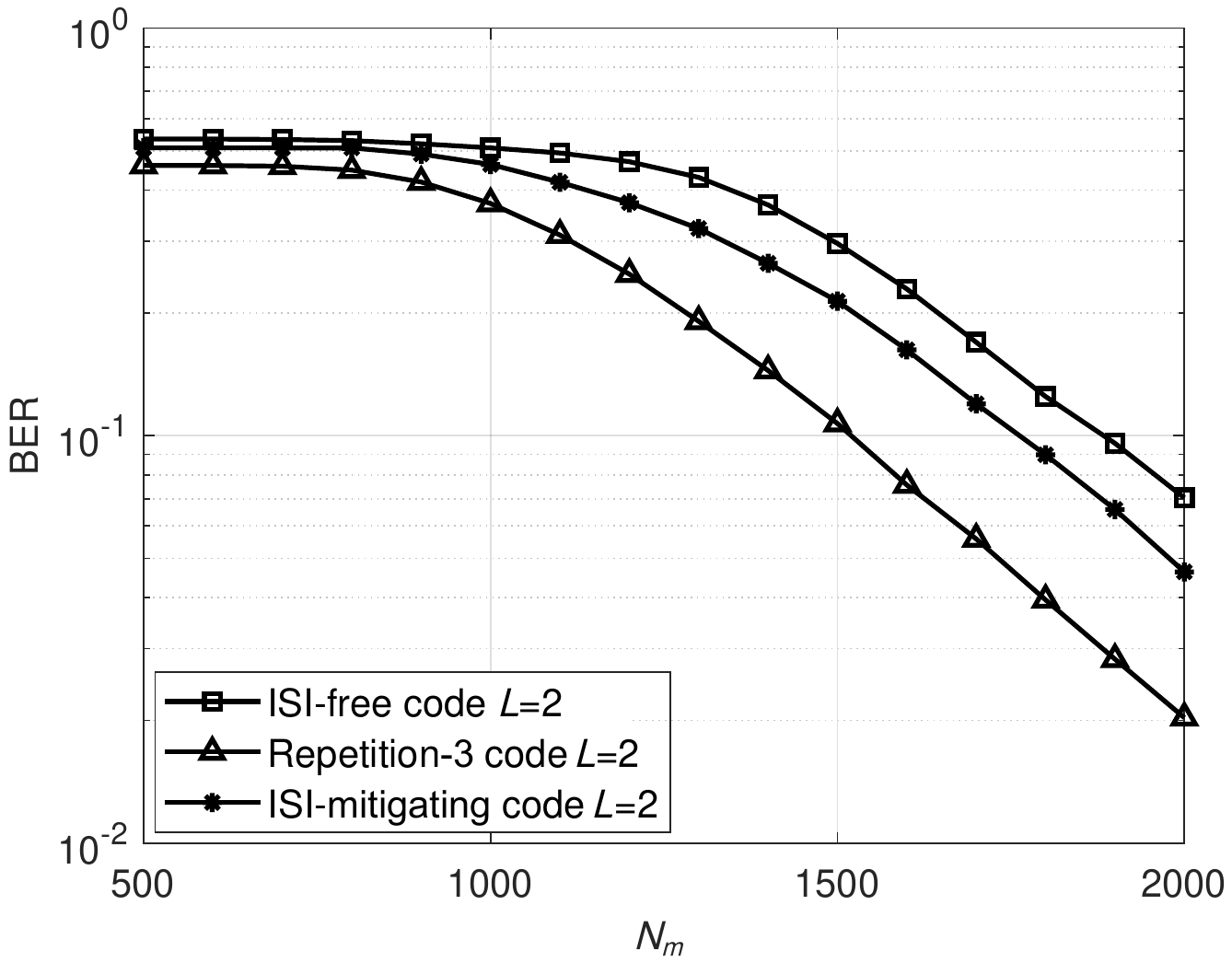}\\
  \caption{Performance comparison of the proposed ISI-mitigating code, ISI-free code, and repetition-3 code varies with the number of transmitted molecules $N_m$.}\label{bervsnm}
\end{figure}

\begin{figure}[!h]
  \centering
  \includegraphics[width=0.45\textwidth]{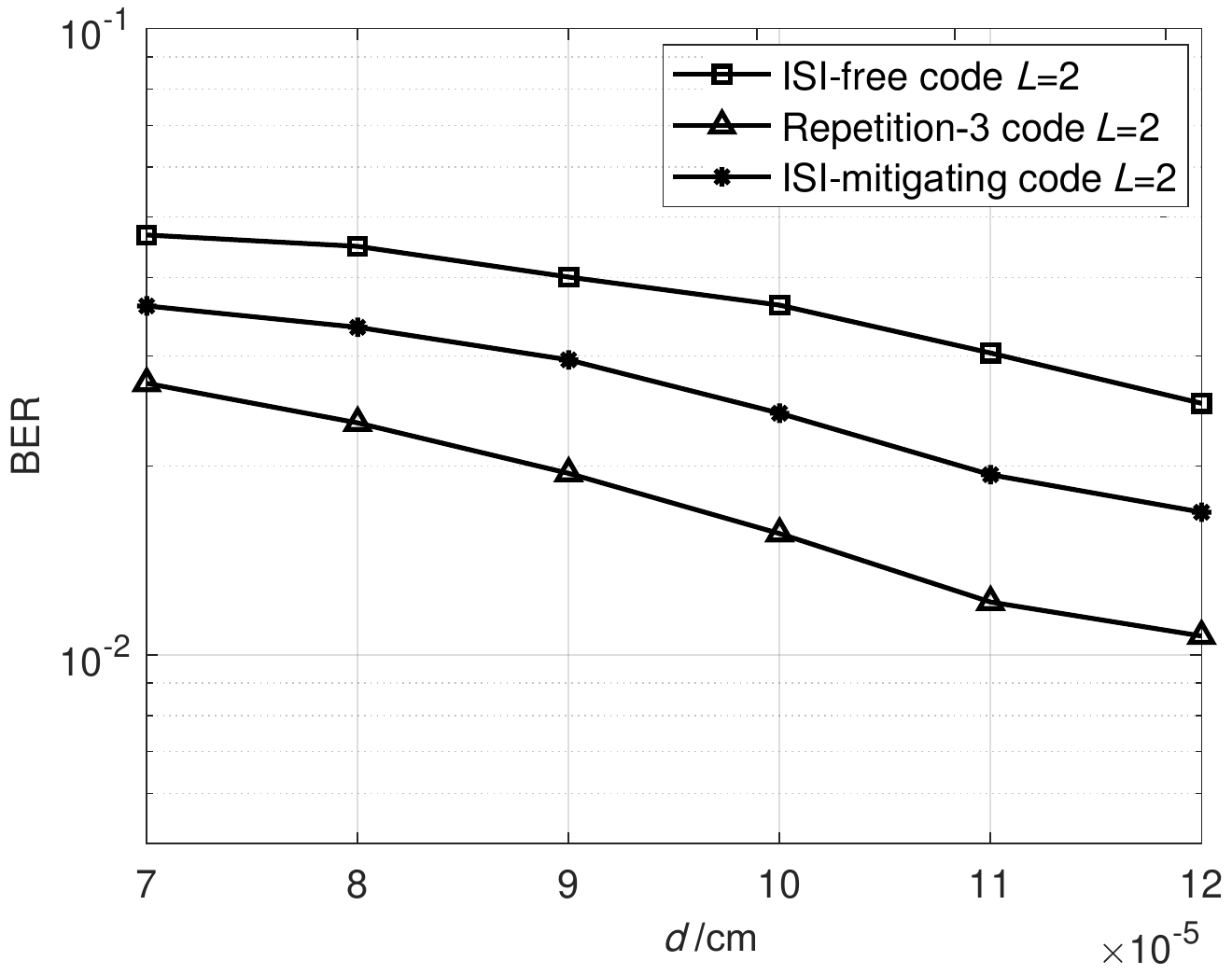}\\
  \caption{Performance comparison of the proposed ISI-mitigating code, ISI-free code, and repetition-3 code varies with the distance between transmitter and receiver.}\label{bervsd}
\end{figure}

\begin{figure}[!h]
  \centering
  \includegraphics[width=0.45\textwidth]{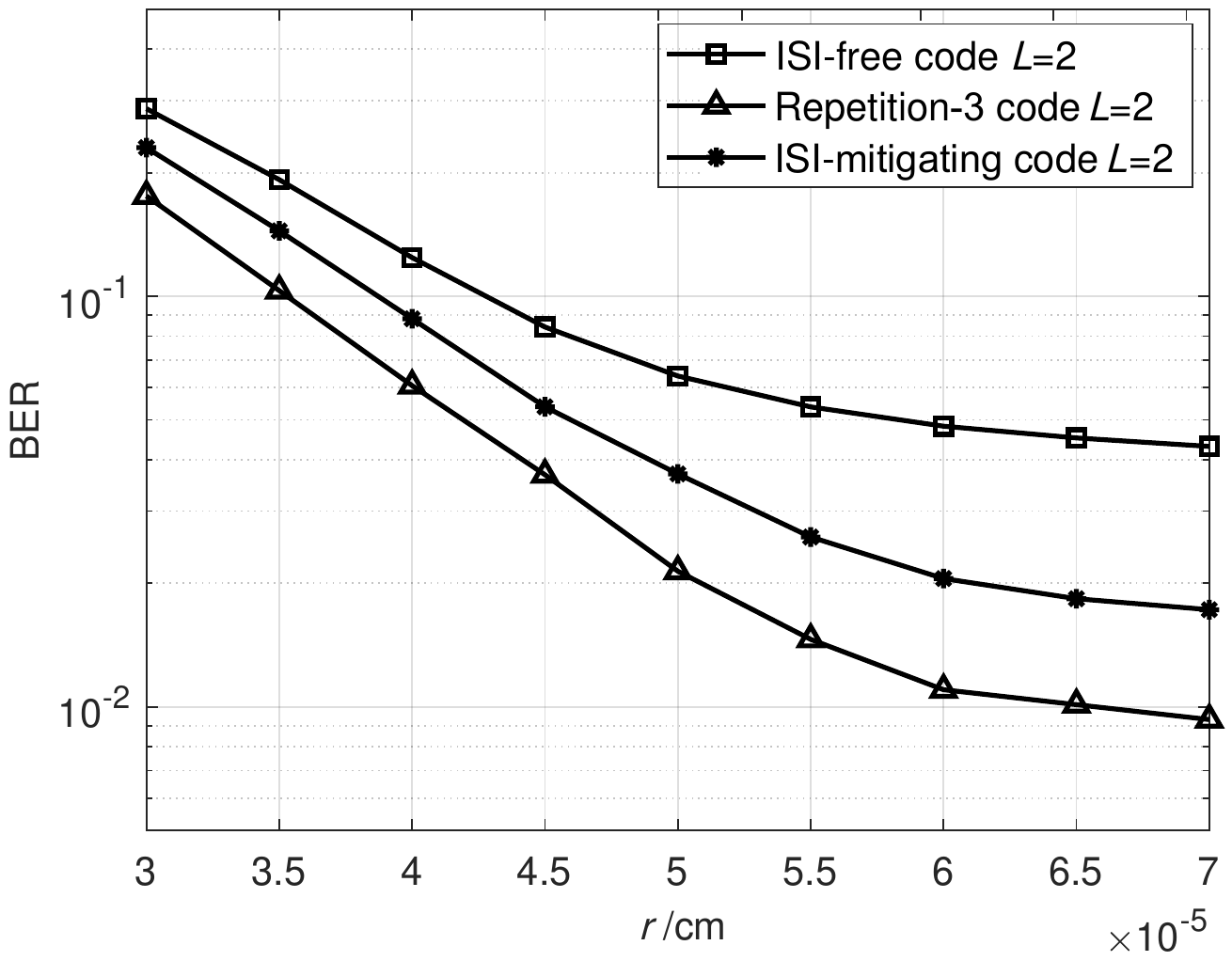}\\
  \caption{Performance comparison of the proposed ISI-mitigating code, ISI-free code, and repetition-3 code varies with the radius of the  receiver.}\label{bervsr}
\end{figure}

The performance of the proposed ISI-mitigating code, ISI-free code, and repetition-3 code against the number of transmitted molecules $N_m$, the distance between transmitter and receiver $d$, and the radius of the receiver $r$ are given in Fig. \ref{bervsnm}, Fig. \ref{bervsd}, and Fig. \ref{bervsr}, respectively. As shown in Fig. \ref{bervsnm}, with the increase of $N_m$, the BER decreases, and the gap between the ISI-mitigating code and the repetition-3 code also decreases, showing the excellent performance of the proposed ISI-mitigating code. In Fig. \ref{bervsd}, the BER decreases with increasing distance between the transmitter and the receiver, though the rate of decrease is small. This is because, with the increase of distance, the number of received ISI molecules decreases. Moreover, the optimal detection threshold is closer to the preset detection threshold, so better performance is achieved. In Fig. \ref{bervsr}, the performance of the proposed ISI-mitigating code, ISI-free code, and repetition-3 code are compared against the radius of the receiver. As shown in Fig. \ref{bervsr}, the BER decreases as the the radius of the receiver increases, since an increase in the radius of the receiver means that more molecules are received. However, the number of ISI molecules also increases, making the slope of the BER decrease with respect to the radius of the receiver. The result in Fig. \ref{bervsr} also indicates the effectiveness of the ISI-mitigating code scheme.

\section{Conclusions}

In this work, we have proposed the ISI-mitigating code technique for mitigating ISI in molecular communication between bionanosensors. The proposed scheme enhances communication reliability and achievable rates under the diffusion channel. Theoretical analysis and simulation results validate that the performance of the ISI-mitigating code is strengthened by avoiding the accumulation of molecules in adjacent time slots. Detailed comparison with existing schemes also shows that our proposed scheme has low complexity, which recommends it as a desirable approach in the context of molecular communication. In the future, channel codes in the molecular communication between bionanosensors with lower complexity and better performance instead of introducing traditional channel codes remains to be studied.

\bibliographystyle{IEEEtran}
\bibliography{references}

\begin{IEEEbiography}[{\includegraphics[width=1in,height=1.25in,clip,keepaspectratio]{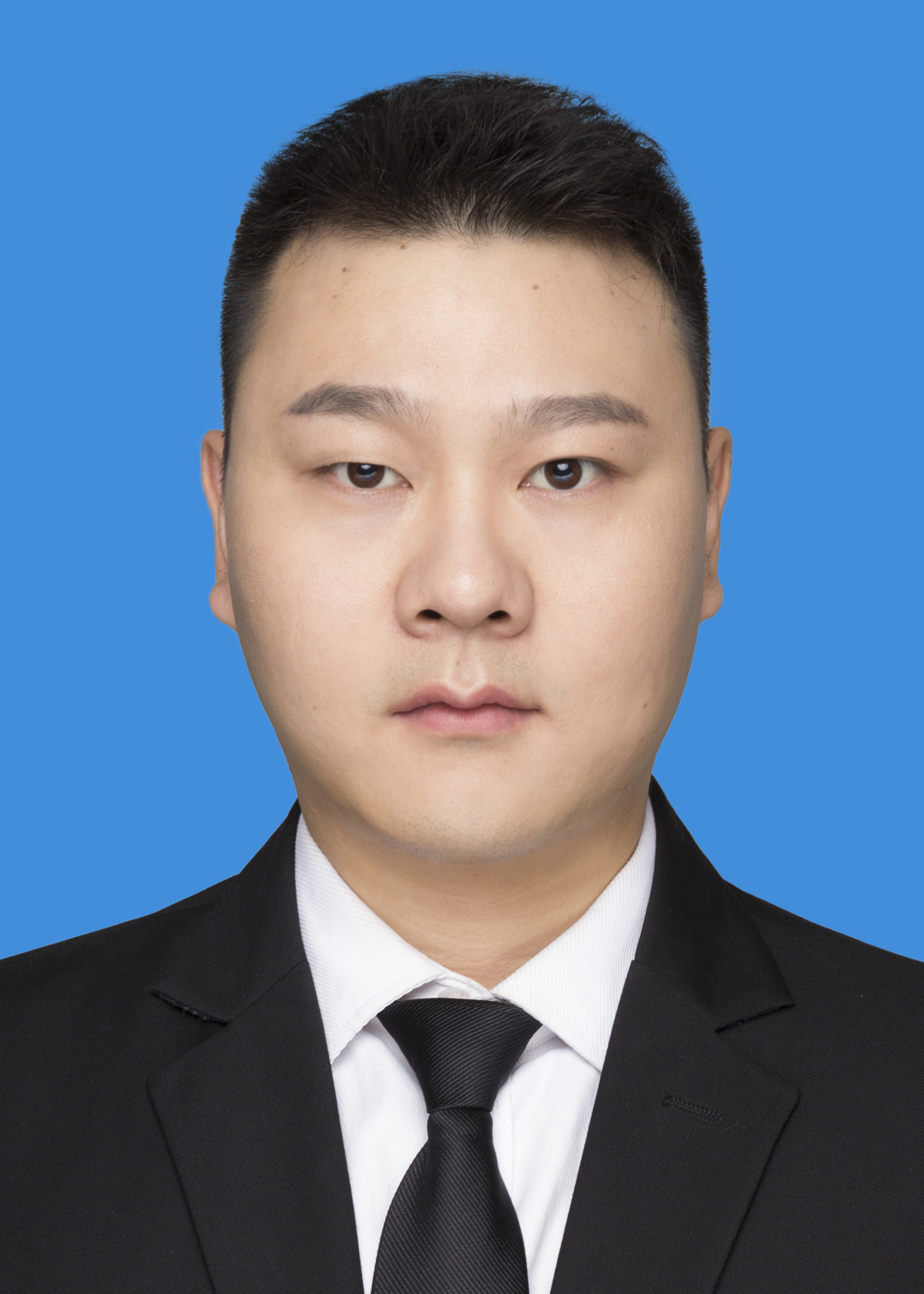}}]{Dongliang Jing} is a lecturer in the the College of Mechanical and Electronic Engineering, Northwest A\&F University, Yangling, China. He received the B.S. degree from Anhui Polytechnic University, Wuhu, China in 2015 and received the Ph.D. degree from the Xidian University, Xian, China in 2022. During Nov. 2019 - Nov. 2020, he was a visiting student for molecular communication in York University, Toronto, ON. Canada, under the supervisor of Andrew W. Eckford. His main research interests include wireless and molecular communications.
\end{IEEEbiography}

\begin{IEEEbiography}[{\includegraphics[width=1in,height=1.25in,clip,keepaspectratio]{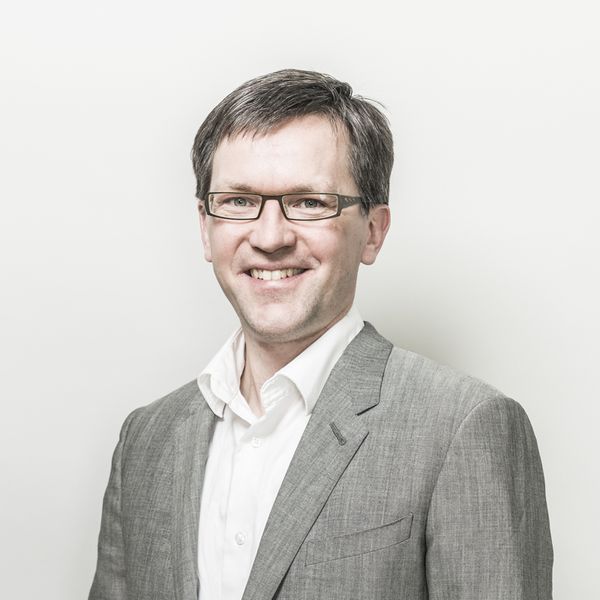}}]{Andrew W. Eckford} is an Associate Professor in the Department of Electrical Engineering and Computer Science at York University, Toronto, Ontario. His research interests include the application of information theory to biology, and the design of communication systems using molecular and biological techniques. His research has been covered in media including The Economist, The Wall Street Journal, and IEEE Spectrum. His research received the 2015 IET Communications Innovation Award, and was a finalist for the 2014 Bell Labs Prize. He is also a co-author of the textbook Molecular Communication, published by Cambridge University Press. Andrew received the B.Eng. degree from the Royal Military College of Canada in 1996, and the M.A.Sc. and Ph.D. degrees from the University of Toronto in 1999 and 2004, respectively, all in Electrical Engineering. Andrew held postdoctoral fellowships at the University of Notre Dame and the University of Toronto, prior to taking up a faculty position at York in 2006. He has held courtesy appointments at the University of Toronto and Case Western Reserve University. In 2018, he was named a Senior Fellow of Massey College, Toronto.
\end{IEEEbiography}
\end{document}